\documentclass[preprint2]{aastex}
\usepackage{natbib}

\newcommand{\nb}{$N$-body } 
\newcommand{\lcdm}{$\Lambda$CDM } 
\newcommand{\n}{\noindent}
\newcommand{\lx}{\hbox{$L_\mathrm{X}$}}
\newcommand{\lxtx}{\hbox{$L_\mathrm{X}$-$T$}} 

\shorttitle{GALAXY CLUSTERS: THE NEED FOR AGN HEATING}
\shortauthors{SHORT \& THOMAS}

\begin{document}

\title{COMBINING SEMI-ANALYTIC MODELS WITH SIMULATIONS OF GALAXY CLUSTERS: \\ THE NEED FOR HEATING FROM ACTIVE GALACTIC NUCLEI} 
\author{C.~J.~SHORT AND P.~A.~THOMAS}
\affil{Astronomy Centre, University of Sussex, Falmer, Brighton, BN1 9QH, United Kingdom}
\email{E-mail: C.Short@sussex.ac.uk}

\begin{abstract}
We present hydrodynamical \nb simulations of clusters of galaxies with feedback taken from
  semi-analytic models of galaxy formation. The advantage of this technique is that the source of feedback in our simulations is a population of galaxies that closely resembles that found in the real
  universe. We demonstrate that, to achieve the high entropy levels found in clusters,
  active galactic nuclei must inject a large fraction of their energy into the
  intergalactic/intracluster media throughout the growth period of the central black hole.
  These simulations reinforce the argument of \citet{BMB08}, who arrived at the same conclusion
  on the basis of purely semi-analytic reasoning.
\end{abstract}

\keywords{hydrodynamics -- methods: N-body simulations -- galaxies: clusters: general -- (galaxies:) cooling flows -- X-rays: galaxies: clusters.}

\section{INTRODUCTION}

X-ray observations of clusters of galaxies allow us to probe the physical properties of
the hot, diffuse intracluster medium (ICM). In the simplest self-similar collapse scenario, where the ICM is
heated solely by gravitational processes (such as adiabatic compression and shocks induced
by supersonic accretion), we expect the X-ray luminosity $\lx$ of clusters to scale with
gas temperature $T$ as $\lx\propto T^2$ (for $T\gtrsim 2$ keV); see
\citet{KAI86}. However, the observed luminosity-temperature relation is much steeper than
predicted, with $\lx\propto T^{2.5-3}$ at $T>2$ keV \citep{MAR98,ARE99,WXF99,EDM02},
becoming steeper still at group scales $T\lesssim 2$ keV \citep{HEP00,XUW00,OSP04}. This
deficit in X-ray luminosity, particularly in low-mass systems, is due to an excess of
entropy \footnote{We define entropy as $S=kT/n_\mathrm{e}^{\gamma-1}$, where $k$ is
  Boltzmann's constant, $n_\mathrm{e}$ is the electron number density and $\gamma=5/3$ is
  the ratio of specific heats for a monoatomic ideal gas.} in cluster cores
\citep{PCN99,LPC00,FJB02}. The source of this excess entropy is likely to be a combination
of non-gravitational cooling and heating processes (see \citealt{VOI05} for a review).

Radiative cooling leads to an increase in the entropy of the ICM since cold, low-entropy
gas condenses to form stars, reducing the gas density and causing higher-entropy gas to
flow inwards to compensate for the loss of pressure support. This leads to a decrease in
X-ray luminosity, breaking the self-similarity of the cluster scaling relations in a way
that resembles observations \citep{BRY00,PTC00,MTK01,MTK02,DKW02,VBB02,WUX02}. However,
hydrodynamical \nb simulations that only include radiative cooling generically predict an
excessive amount of cold gas that can form stars (e.g. \citealt{SUO98,LBK00,YSS02});
typically the fraction, $f_*$, of baryons in stars lies in the range $0.3-0.5$ in
simulated clusters, in clear conflict with the observed value of $f_*\lesssim 0.1$
\citep{BPB01,LMS03,BMBE08}. This highlights the need for extra heating (\emph{feedback})
from astrophysical sources to regulate the cooling flow and quench star formation.

Non-gravitational heating, occurring before or during gravitational collapse, raises the
entropy of the ICM, preventing gas from reaching high densities in central cluster regions
and reducing its X-ray emissivity. This effect will be greater in lower-mass systems,
leading to a steepening of the $\lx$-$T$ relation as desired. The most obvious sources of
non-gravitational heating are Type II supernovae (SNe) and active galactic nuclei (AGNs). In the simplest heating models, the
energy released by these phenomena is assumed to be injected impulsively into the gas at
high redshift. Although not well motivated physically, these so-called \emph{preheating}
models have proved capable of reproducing the observed slope and normalisation of cluster scaling relations when incorporated in simulations (e.g. \citealt{BEM01,BRM01,MTK02,BGW02,TBS03,BFK05}). However, the preheating scenario suffers from several problems. For example, the predicted scatter about the mean $\lx$-$T$ relation is much smaller than observed and large isentropic cores are generated in low-mass systems that are not seen (e.g. \citealt{PSF03,PAP06}).

Recently, attention has shifted to more realistic models which attempt to couple cooling,
star formation and black hole growth with feedback from SNe and AGNs. There has been
considerable effort to include such processes in hydrodynamical simulations of galaxy
groups and clusters. However, an explicit treatment is unfeasible since star formation,
black hole growth and associated feedback all occur on scales much smaller than can be
resolved with present computational resources. The only option is to develop
phenomenological prescriptions and assess their validity by comparing the properties of
simulated clusters with as much observational data as possible.

Various theoretical models of stellar feedback have been proposed and can be grouped in
two basic categories: \emph{thermal}, where the available supernova energy is used to
raise the temperature of neighbouring gas particles, and \emph{kinetic}, where
neighbouring particles are given a velocity `kick'. Cosmological simulations with
radiative cooling, star formation and supernova feedback yield ICM profiles and scaling
relations that are in reasonable agreement with observations, for both thermal
(e.g. \citealt{VAL03,KAY04,KTJ04,KDA07,NKV07}) and kinetic
(e.g. \citealt{BMS04,BFK05,RSP06}) models. In particular, the mean $\lx$-$T$ relation is generally well reproduced on cluster scales $T\gtrsim 2$ keV. However, X-ray luminosities are found to be
substantially larger than the observed values on group scales. Other problems are that
observed baryon fractions are typically smaller than in simulations
(e.g. \citealt{KNV05,EDB06}), and star formation is still too efficient, indicating that
additional feedback mechanisms are required to offset radiative cooling.

The favoured candidate is the gravitational energy liberated by the accretion of gas onto
central supermassive black holes within galaxies. This can be extremely large, of the
order $10^{62}(M_\mathrm{BH}/10^{9}M_{\sun})$ erg, where $M_\mathrm{BH}$ is the black hole
mass. Analytical calculations suggest that the $\lx$-$T$ relation of groups can be
accounted for if $\sim 1\%$ of the energy released by AGNs is coupled to the surrounding
ICM (e.g. \citealt{CLM02}). The precise details of how this energy is transferred to the
ICM are not well understood at present, but it appears there are two major channels via
which black holes interact with their surroundings (see \citealt{MCN07} for a review).

At high redshift, mergers of gas-rich galaxies occur frequently and are expected to funnel
copious amounts of cold gas towards galactic centres, leading to high black hole accretion
rates and radiating enough energy to support the luminosities of powerful
quasars. Quasar-induced outflows have been observationally confirmed in a number of cases
(e.g. \citealt{CBG03,CKG03,PRK03}), and demonstrated in high-resolution simulations of
galaxy mergers. In such simulations it is usually assumed that a small fraction of the
bolometric luminosity can couple thermally to the surrounding gas. This approach has been
shown to successfully reproduce the observed black hole-bulge mass relation
(\citealt{DSH05,RHC06}) and explain the red colours \citep{SDH05b}, X-ray haloes
\citep{CDH06} and fundamental plane relation (\citealt{RCH06}) of massive elliptical
galaxies. \citet{DCS08} recently extended this work to fully cosmological simulations,
obtaining similarly encouraging results.

There is another mode of AGN feedback which is not related to quasar activity or triggered
by galaxy mergers. Evidence for this can be seen, for example, in brightest cluster
galaxies (BCGs) which contain X-ray cavities filled with relativistic plasma
(e.g. \citealt{BSM01,BRM04,MNW05,FST06}). These radio-loud X-ray depressions, referred to
as \emph{bubbles}, are thought to be inflated by relativistic jets launched from the
central black hole. Bubbles may rise buoyantly, removing some of the central cool gas and
allowing it to mix with hotter gas in the outer regions of groups and clusters. Together
with the accompanying mechanical heating, this can constitute an efficient mechanism for
suppressing cooling flows. Simulations of idealised clusters, performed with
hydrodynamical mesh codes, suggest that this is indeed the case
(e.g. \citealt{CBK01,QBB01,RUB02,BKC02,BRU03,DBT04}). 

The first attempt to implement a
self-consistent model of black hole growth and heating by AGN-driven bubbles in
cosmological simulations was made by \citet{SIS06}. They found that bubble injection can
substantially affect the properties of the ICM, especially in massive, relaxed systems at
late times. In particular, bubbles are able to efficiently heat central cluster gas,
reducing the amount of cold baryons and star formation in the central cD
galaxy. Furthermore, the gas density is reduced and the temperature is increased out to
radii $\sim 300 h^{-1}$ kpc, leading to a decline in X-ray luminosity and establishing a
flat entropy profile in central regions. Such trends are precisely what is required to
reconcile simulations of galaxy clusters with observations.

Following this work, \citet{SSD07} have formulated a unified model of AGN feedback,
incorporating both `quasar' and `radio' modes, as well as star formation and feedback from
SNe. Motivated by the properties of X-ray binaries (e.g. \citealt{FCT99,GFP03}), the
transition between the two states is assumed to be governed by an accretion rate threshold
$\chi_\mathrm{radio}=\dot{M}_\mathrm{BH}/\dot{M}_\mathrm{Edd}$, where
$\dot{M}_\mathrm{Edd}$ is the Eddington accretion rate. Quasar-like feedback occurs for
accretion rates greater than $\chi_\mathrm{radio}$, while mechanical bubble heating takes
place otherwise. This model has recently been used by \citet{PSS08} to investigate the
$\lx$-$T$ relation and gas fractions of galaxy groups and clusters in cosmological
simulations. They demonstrated that AGN feedback reduces the X-ray luminosities of groups
and poor clusters more than rich clusters, resulting in a steepening of the $\lx$-$T$
relation on group scales. In fact, the X-ray properties of their simulated objects are in excellent agreement with observational data on all mass scales. However, since their sample size is quite small (21 objects), it is unclear whether their model can generate a realistic population of cool core (CC) and non-cool core (NCC) systems and thus explain the observed scatter about the mean $\lx$-$T$ relation. The gas fraction in groups and poor
clusters was shown to decrease significantly with the inclusion of AGN heating (even
though fewer baryons were converted into stars), because gas is driven from central
regions to cluster outskirts. In more massive systems, the main effect of AGN feedback is
to lower the central gas density and substantially reduce the number of stars
formed. However, even with stellar and AGN feedback, the stellar fraction within the
virial radii of their simulated objects appears larger than observations suggest, implying
that the cooling flow problem has not been fully resolved.

In this paper we pursue a different, but complementary, approach to the theoretical study
of groups and clusters of galaxies. Instead of undertaking fully self-consistent
hydrodynamical simulations, we investigate what current semi-analytic models (SAMs) of
galaxy formation predict for the thermodynamical properties of the ICM. Our goal is to
extend the predictive power of these models, thus providing additional constraints on, and
insights into, the physics of galaxy formation. We hope that our work will be useful for
guiding the development of future SAMs that can simultaneously account for the properties
of both the galaxy distribution and the ICM.

The basis of a SAM is a set of dark matter halo merger trees, usually obtained from a
high-resolution \nb simulation. The behaviour of baryonic matter within these dark haloes
is modelled using analytical `recipes' to capture the essential physical processes
involved in the formation and evolution of galaxies: gas cooling, star formation, black
hole growth, feedback, galaxy dynamics, galaxy mergers, etc. These recipes typically
contain a number of adjustable parameters, which are tuned to attain the best possible
match to selected observational properties of the galaxy distribution. This semi-analytic
approach has proved largely successful, reproducing many key properties of real galaxies
such as luminosities, colours, star formation rates, the Tully-Fisher relation and the
black hole-bulge mass relation
(e.g. \citealt{KCD99,CLB00,KAH00,BBF03,DKW04,GDS04,BBM06,CSW06,CDD06,MFG06,MFT07,FBM08}).

The essence of our method is to generate a semi-analytic galaxy catalogue for a
cosmological \nb simulation, compute the energy released by SNe and the AGN in each model
galaxy, and inject this energy at the appropriate position and time into the baryonic
component of a non-radiative hydrodynamical simulation that has the \emph{same}
distribution of dissipationless dark matter. In this way we can track the effect of energy
feedback from semi-analytic galaxies on the intracluster gas. A similar technique has
already been used to study the metal enrichment of the ICM \citep{COR06,CTT08}. 

There are several benefits of our hybrid approach. Firstly, feedback is guaranteed to originate from a galaxy population whose observational properties agree well with those of real galaxies. This is generally not the case in self-consistent hydrodynamical simulations. Secondly, we only need a single dark matter
simulation to construct a semi-analytic galaxy catalogue. Since \nb simulations are much
less computationally demanding than hydrodynamical simulations, we can, in principle,
attain significantly greater resolution for the collisionless component. With a
high-resolution dark matter simulation we can construct a comprehensive galaxy catalogue
and thus a detailed model for feedback from galaxies. Thirdly, the energy transferred to
the ICM by SNe and AGNs can be calculated directly from the semi-analytic galaxy catalogue,
before coupling the SAM to a hydrodynamical simulation. By avoiding the need to include an
explicit sub-grid model for star formation, black hole growth and associated feedback
processes, the hydrodynamical simulations we use to track the injection of energy into the
ICM require considerably less computational effort. Finally, as we shall see, a lower
resolution can be used for the gas than the dark matter in our hydrodynamical
simulations. As a result, our technique is readily applicable to large cosmological
volumes, allowing massive clusters to be simulated.

The layout of this paper is as follows. In Section \ref{sec:method} we discuss our
numerical method in detail, describing the simulations used, the relevant components of
the SAM and our feedback implementation. We use our hybrid approach to investigate the
bulk properties of the ICM in Section \ref{sec:results}, comparing our results with a
selection of observational data. A resolution test is presented in Section
\ref{sec:restest} to demonstrate the robustness of our results. We summarise our results
and conclude in Section \ref{sec:conc}.

\section{THE NUMERICAL MODEL}
\label{sec:method}

Our hybrid technique for studying the effect of galaxy feedback on intracluster gas
consists of three distinct components: an underlying dark matter simulation, a
semi-analytic galaxy catalogue built on the halo merger trees of this simulation, and a
hydrodynamical simulation to track the energy injection from model galaxies. In this
section we provide a detailed description of each part of the modelling process.

\subsection{Dark matter simulations}
\label{sec:dmsim}

The cosmological model we adopt is a spatially-flat \lcdm model with cosmological
parameters $\Omega_\mathrm{m,0}=0.25$, $\Omega_{\Lambda,0}=0.75$, $h=0.73$,
$n_\mathrm{s}=1$ and $\sigma_{8,0}=0.9$. Here $\Omega_\mathrm{m,0}$ and
$\Omega_{\Lambda,0}$ are the total matter and dark energy density parameters, $h$ is the
Hubble parameter $H_0$ in units of $100$ km s$^{-1}$ Mpc$^{-1}$, $n_\mathrm{s}$ is the
spectral index of primordial density perturbations, and $\sigma_{8,0}$ is the rms linear
density fluctuation within a sphere of radius $8h^{-1}$ Mpc. The subscript $0$ signifies
the value of a quantity at the present day. While there is some tension between our chosen
parameter values (particularly $n_{\rm s}$ and $\sigma_{8,0}$) and those derived from the
five-year Wilkinson Microwave Anisotropy Probe (WMAP) data \citep{DKN09}, we have
deliberately used the same cosmological parameters as the Millennium simulation
\citep{SWJ05} since we eventually hope to apply our technique to the full Millennium
volume, using the publicly-available Millennium semi-analytic galaxy catalogues
\citep{LEM06} as input.

Initial conditions were created at a redshift $z_\mathrm{i}=127$ by displacing particles
from a glass-like distribution, so as to form a random realisation of a density field with
a \lcdm linear power spectrum obtained from CMBFAST \citep{SEZ96}. The amplitudes and
phases of the initial density perturbations were chosen to be the same as those of the
Millennium simulation initial conditions. We generated initial conditions for two cubic
simulation volumes: one with a comoving side length $L=62.5 h^{-1}$ Mpc and
$N_\mathrm{DM}=270^3$ dark matter particles, and the other with $L=125 h^{-1}$ Mpc and
$N_\mathrm{DM}=540^3$. The mass of a dark matter particle is then
$m_\mathrm{DM}=8.61\times 10^{8}h^{-1}M_{\sun}$, as in the Millennium simulation. The smaller
of these simulations is for the purposes of initial model discrimination, while the larger
simulation is to demonstrate our best model in a more cosmologically interesting volume.

The massively parallel tree $N$-body/SPH code GADGET-2 \citep{SPR05} was then used to
evolve the initial conditions to $z=0$, with full particle data stored at the $64$ output
redshifts of the Millennium simulation: $z_{63}=127$, $z_{62}=80$, $z_{61}=50$,
$z_{60}=30$ and $\log_{10}(1+z_n)=n(n+35)/4200$, $0\leq n<60$. The Plummer-equivalent
gravitational softening length was fixed at $\epsilon=40h^{-1}$ kpc in comoving
coordinates until $z=3$, then fixed in physical coordinates thereafter. Note that
gravitational forces were softened on a fixed comoving scale of $\epsilon=5h^{-1}$ kpc in
the Millennium simulation, a factor of two smaller than our $z=0$ softening length. We
chose a larger softening scale for two reasons. Firstly, by comparing the number of dark
matter substructures (see below) formed in simulations with $\epsilon=5h^{-1}$ kpc and
$\epsilon=10h^{-1}$ kpc, we found that the more aggressive softening scheme yields fewer
low-mass objects because of two-body heating effects. This means we can construct a more
detailed semi-analytic galaxy catalogue by setting $\epsilon=10h^{-1}$ kpc. Secondly, we
will be incorporating gas particles into our dark matter simulations and it has been shown
that an optimal choice for the softening length in hydrodynamical simulations is
approximately $4\%$ of the mean inter-particle spacing, corresponding to $\epsilon\approx
10h^{-1}$ kpc in our case \citep{THC92,BDM06}.

\subsubsection{Dark matter haloes, substructure and merger tree construction}

Dark matter haloes are identified as virialised particle groups within the simulations
using the friends-of-friends (FOF) algorithm. We adopt a standard FOF linking length of
$20\%$ of the mean particle separation \citep{DEF85} and only save groups that contain at
least $20$ particles, so that the minimum halo mass is $1.7\times 10^{10}h^{-1}M_{\sun}$. FOF
group catalogues are produced on the fly and in parallel by the simulation code. An
improved version of the SUBFIND algorithm \citep{SWT01} is then applied in post-processing
to these group catalogues to find gravitationally-bound dark matter substructures orbiting
within the FOF haloes.

To compute a virial mass estimate for each FOF halo, a sphere is grown about the minimum
of the gravitational potential within the group until the mean overdensity enclosed
reaches $\Delta\rho_\mathrm{cr}(z)$, where $\Delta$ is the desired density contrast,
$\rho_\mathrm{cr}(z)=3H_0^2E(z)^2/8\pi G$ is the critical density and $E(z)^2=\Omega_{\rm
  m,0}(1+z)^3+\Omega_{\Lambda,0}$ in a \lcdm cosmological model. The mass enclosed within
the sphere, $M_\Delta$, is then related to the sphere radius, $r_\Delta$, and the circular
velocity, $v_\Delta$, at this radius by

\begin{equation}
M_\Delta=\frac{4\pi}{3} r_\Delta^3\Delta\rho_\mathrm{cr}(z)=\left[\frac{3}{4\pi
    G^3\Delta\rho_\mathrm{cr}(z)}\right]^{1/2}v_\Delta^3.
\end{equation}

\n Following \citet{CSW06}, a density contrast $\Delta=200$ is used to define the virial
mass, radius and velocity of a halo throughout this paper. 

Once all haloes and subhaloes have been identified for each simulation output, we
construct \emph{merger trees} that describe how haloes grow as the universe evolves. This
is done by exploiting the fact that each halo will have a unique descendant in a
hierarchical scenario of structure formation; see \citet{SWJ05} for further details. Each
individual merger tree then contains the full formation history of a given halo at $z=0$.

\subsection{The semi-analytic model of galaxy formation}

Dark matter halo merger trees form the backbone of modern hierarchical models of galaxy
formation. We have generated galaxy catalogues for our two dark matter simulations by
applying a SAM to to the merger trees. The SAM we use is the highly successful Munich
L-Galaxies model described by \citet{DLB07}. We adopt the same set of model parameters as
\citet{DLB07} since these parameters were used to produce the publicly-available
Millennium galaxy catalogue. As in the full Millennium catalogue, galaxy and host halo
properties are stored at the same $64$ redshift values as the simulation outputs. We now
review the components of this model that are relevant for our work; for a full description
of L-Galaxies we refer the reader to \citet{CSW06} and \citet{DLB07}.

\subsubsection{Star formation and supernova feedback}

The infall of gas into the potential well of a dark halo causes the gas to be shock heated
to the halo virial temperature. At late times and in massive systems, this gas is added to
a quasi-static hot atmosphere that extends roughly to the virial radius of the dark
halo. Gas from the inner regions of this atmosphere can then cool and accrete onto a
central cold gas disk. At early times and in lower mass systems, the post-shock cooling
time is sufficiently short that a quasi-static halo cannot form. Instead the shocked gas
rapidly cools and settles onto the cold disk.

Once gas in the cold disk exceeds a critical surface density, it can collapse and form
stars \citep{KEN89}. Massive stars rapidly complete their life cycle and explode as Type
II SNe, injecting gas, metals and energy into the surrounding medium, reheating cold disk
gas and possibly ejecting gas from the quasi-static hot halo. For a given mass $\Delta
m_*$ of stars formed over some finite time interval, the amount of energy released by Type
II SNe that is available for heating gas is approximated by

\begin{equation}
\Delta E_\mathrm{SN}=\frac{1}{2}\epsilon_\mathrm{halo}\Delta m_* v_\mathrm{SN}^2,
\end{equation}

\n where $v_\mathrm{SN}=630$ km s$^{-1}$, based on a standard stellar initial mass
function (IMF), and $\epsilon_\mathrm{halo}$ is an efficiency parameter. The published
value for this parameter is $\epsilon_\mathrm{halo}=0.35$ \citep{CSW06}. We note that the
mass actually locked-up in stars is given by $\Delta M_*=(1-R)\Delta m_*$, the rest is
assumed to be instantaneously returned to the cold disk. Here $R$ is the recycle fraction,
which is assigned the value $R=0.43$ in accordance with the Chabrier IMF \citep{CHA03}
employed by \citet{DLB07}.

The mass of cold gas reheated by SNe is modelled as

\begin{equation}
\Delta m_\mathrm{reheated}=\epsilon_\mathrm{disk}\Delta m_*,
\end{equation}

\n where $\epsilon_\mathrm{disk}$ is a parameter that is set to
$\epsilon_\mathrm{disk}=3.5$, motivated by the observational work of \citet{MAR99}. If the
reheated gas were added to the hot halo without changing its specific energy, its total
thermal energy would increase by

\begin{equation}
\Delta E_\mathrm{hot}=\frac{1}{2}\Delta m_\mathrm{reheated} v_\mathrm{vir}^2.
\end{equation}

\n If there is any excess energy after reheating, i.e. $\Delta E_\mathrm{SN}>\Delta E_{\rm
  hot}$, then it is assumed that a mass

\begin{equation}
\label{eq:mej}
\Delta m_\mathrm{ejected}=\left[\epsilon_\mathrm{halo}\left(\frac{v_\mathrm{SN}}{v_{\rm
      vir}}\right)^2-\epsilon_\mathrm{disk}\right]\Delta m_*
\end{equation}

\n of hot gas is ejected from the halo into an external `reservoir'. When $\Delta E_{\rm
  SN}<\Delta E_\mathrm{hot}$, there is insufficient energy to eject any gas out of the
halo and $\Delta m_\mathrm{ejected}$ is set to zero. Note that the reheated mass is not
reduced in this case. It follows from equation (\ref{eq:mej}) that no hot gas can be
expelled if
$v_\mathrm{vir}>(\epsilon_\mathrm{halo}/\epsilon_\mathrm{disk})^{1/2}v_\mathrm{SN}\approx
200$ km s$^{-1}$, whereas the entire hot halo can be ejected for small $v_\mathrm{vir}$.

\subsubsection{Black hole growth and cooling flow suppression}

The growth of supermassive black holes in L-Galaxies is driven primarily by quasar mode
accretion, as a result of galaxy mergers. Black holes can grow either by merging with each
other or by the accretion of cold disk gas. The coalescence of black holes is modelled
simply by taking the sum of the progenitor black hole masses. The gas mass $\Delta M_{\rm
  BH,Q}$ accreted during a merger of galaxies with respective masses $M_\mathrm{sat}$ and
$M_\mathrm{central}$ is

\begin{equation}
\label{eq:quasar}
\Delta M_\mathrm{BH,Q}=\frac{f_\mathrm{BH}(M_\mathrm{sat}/M_\mathrm{central})M_{\rm
    cold}}{1+(280\;{\rm km\;s}^{-1}/v_\mathrm{vir})^2},
\end{equation}

\n where $M_\mathrm{cold}$ is the total cold gas mass present and the constant $f_{\rm
  BH}=0.03$ is chosen to reproduce the observed local black hole-bulge mass
relation. Black hole accretion occurs during both minor ($M_\mathrm{sat}\ll
M_\mathrm{central}$) and major ($M_\mathrm{sat}\approx M_\mathrm{central}$) mergers,
although the efficiency in the former case is reduced by the
$M_\mathrm{sat}/M_\mathrm{central}$ term.

Once a static hot halo has formed around the host galaxy of a black hole, it is assumed
that there is also continual and quiescent accretion onto the central black hole directly
from the hot phase. The growth rate $\dot{M}_\mathrm{BH,R}$ of the black hole in this
radio mode is described by:

\begin{equation}
\label{eq:radio}
\dot{M}_\mathrm{BH,R}=\kappa_\mathrm{AGN}\left(\frac{M_{\rm
    BH}}{10^8M_{\sun}}\right)\left(\frac{f_\mathrm{hot}}{0.1}\right)\left(\frac{v_{\rm
    vir}}{200\;{\rm km\;s}^{-1}}\right)^3,
\end{equation}

\n where $f_\mathrm{hot}$ is the fraction of the total halo mass in the form of hot gas
and $\kappa_\mathrm{AGN}$ is a free parameter controlling the efficiency of accretion. A
value of $\kappa_\mathrm{AGN}=7.5\times 10^{-6}M_{\sun}$ yr$^{-1}$ is found to reproduce the
turnover at the bright end of the galaxy luminosity function \citep{DLB07}. This simple
phenomenological model may represent the accretion of cold gas clouds, or Bondi-Hoyle
accretion from hot gas that fills the space between these clouds \citep{CSW06}.

The mechanical heating associated with radio mode accretion is assumed to impede, or even
prevent, the cooling flow in central regions of the halo. More specifically, the radiated
luminosity is taken to be

\begin{equation}
L_\mathrm{BH,R}=\epsilon_\mathrm{r}\dot{M}_\mathrm{BH,R}c^2,
\end{equation}

\n where $c$ is the speed of light and $\epsilon_\mathrm{r}$ describes how efficiently
matter can be converted to energy near the event horizon. The standard value
$\epsilon_{\rm r}=0.1$ is adopted, which is an approximate value for radiatively efficient
accretion onto a non-rapidly spinning black hole \citep{SHS73}. This injection of energy
reduces the rate at which hot gas is able to cool onto the cold disk from
$\dot{M}_\mathrm{cool}$ to $\dot{M}_\mathrm{cool}-2L_\mathrm{BH}/v_\mathrm{vir}^2$, with
the restriction that the cooling rate remains non-negative. The effectiveness of radio
mode AGN feedback in suppressing cooling flows is greatest at late times and for large
black hole masses, precisely what is required to reproduce the luminosities and colours of
low-redshift, bright galaxies.

\subsection{Hydrodynamical simulations}

To explore the effect of energy feedback from galaxies on the properties of the ICM, we
couple the L-Galaxies SAM to hydrodynamical simulations. The initial conditions for these
simulations are the same as those used for the two dark matter simulations described
above, but we add gas particles with \emph{zero} gravitational mass. The gas particles
then act purely as `tracers' of the dark matter. In this way we ensure that the dark
matter distribution remains unaffected by the inclusion of baryons, so that the halo
merger trees used to generate the semi-analytic galaxy catalogues are the same.

Recent work has shown that the dissipative nature of the baryon fluid can have
  an influence on the structure of haloes \citep{SRE09,RSH09,PTS09} and the growth of
  merger trees \citep{SDD08}.  However, all semi-analytic modelling to date ignores such
  complications and uses merger trees based solely on the dark matter distribution. We are
  forced to follow this route so that we can use SAM input into our simulations.  For the
  purposes of the results presented in this paper, the dark matter approximation is
  unlikely to make any significant difference to our conclusions.

The number of gas particles that we add is $N_\mathrm{gas}=100^3$ for our small
($L=62.5h^{-1}$ Mpc) simulation volume, and $N_\mathrm{gas}=200^3$ for our
larger ($L=125h^{-1}$ Mpc) volume. We choose to include gas at (approximately)
the lower resolution of the Millennium Gas simulations \footnote{The Millennium
  Gas simulations (Pearce et al., in preparation) are a suite of large
  hydrodynamical simulations, all having the same volume as the Millennium
  simulation ($L=500h^{-1}$ Mpc) and utilising the same amplitudes and phases
  for the initial perturbations. The cosmological parameters are also identical,
  except that the present baryon density parameter $\Omega_\mathrm{b,0}=0.045$
  to reflect the inclusion of gas. The simulations contain
  $N_\mathrm{DM}=1000^3/2$ dark matter and $N_{\rm gas}=1000^3/2$ gas particles,
  with respective masses $1.42\times10^{10}h^{-1}M_{\sun}$, and $3.12\times
  10^9h^{-1}M_{\sun}$.}  so that out technique remains computationally feasible
when applied to the full Millennium volume in future work. In Section
\ref{sec:restest} we conduct a resolution test to demonstrate that increasing
the resolution of the gas component relative to the dark matter has a negligible
effect on our results.

The initial conditions are evolved from $z_\mathrm{i}=127$ to $z=0$ with a modified
version of GADGET-2, designed to allow for gas particles with zero gravitational
mass. Whenever a SPH calculation is to be done, we assume $\Omega_\mathrm{b,0}=0.045$ and
assign the gas particles their corresponding true mass: $m_\mathrm{gas}=3.05\times 10^9
h^{-1}M_{\sun}$. This guarantees that gas properties such as density and entropy are computed
correctly. In addition, gas particles are also given their true mass for simulation data
dumps, with the mass of the dark matter particles accordingly reduced to
$(1-f_\mathrm{b})m_{\rm DM}=7.05\times 10^8 h^{-1}M_{\sun}$, where
$f_\mathrm{b}=\Omega_\mathrm{b,0}/\Omega_\mathrm{m,0}$ is the mean cosmic baryon fraction.

In all of our hydrodynamical simulations we choose to neglect gas cooling
  processes since radiative cooling, star formation, black hole growth and associated
  feedback are incorporated in the SAM. Note, however, that gas particles are still
  converted to dissipationless 'star' particles as dictated by the SAM, following the procedure outlined in Section \ref{sec:starform} below.

We appreciate that the distribution and cooling of gas in haloes is treated in rather a
simplistic manner in existing SAMs. However, we emphasise that, to the best of our knowledge, this is the \emph{first} time the effect of energy feedback from galaxies on the ICM has been investigated using a SAM coupled to hydrodynamical simulations. As such, we believe it is sensible to begin with the simplest possible model where cooling is driven entirely by the SAM. In any case, the inclusion of gas cooling could only lead to a decrease in the entropy of intracluster gas, which would reinforce our conclusions with regard to the high degree of feedback that is required to explain the high entropy levels found in clusters. 

A more self-consistent approach would be to include radiative cooling in our hydrodynamical simulations and use the gas distribution to inform the SAM. This extension of the semi-analytic technique would require the simulation and the SAM to be coupled in such a way that both can be undertaken simultaneously. Extensive testing would be necessary to ensure that such a model was as successful as current SAMs in reproducing observed galaxy properties. Such a scheme is a long-term goal of our work but is beyond the scope of this paper.

The semi-analytic galaxy catalogues that we have generated contain the positions and
properties of all model galaxies at $64$ redshift values, corresponding to the output
times of our simulations. We have further modified GADGET-2 so that, once an output
redshift is reached, temporary `galaxy' particles are introduced at the appropriate
locations in the simulation volume. These galaxy particles have a set of associated
properties, such as the change in stellar mass and energy released by SNe/AGNs since the
last output, which are calculated from the SAM galaxy catalogues prior to the
simulation. We use this information to form stars and heat gas in the neighbourhood of
each galaxy, as described in detail in the next section. Following star formation and the
injection of energy, the galaxy particles are removed and the simulation progresses until
the next output time, when the process is repeated.

Note that the properties of the ICM should not be affected by the frequency with which energy is injected as the time interval between our chosen $64$ model outputs is always less than the galaxy halo dynamical time $t_{\rm dyn}=r_{\rm vir}/v_{\rm vir}=0.1/H(z)$. We have verified that increasing the temporal resolution has a negligible effect on our results.

Cluster catalogues are constructed at $z=0$ for our simulations using a procedure similar
to that employed by \citet{MTK02}. Briefly, groups of dark matter particles are identified
with the FOF algorithm, setting the linking length to be $10\%$ of the mean inter-particle
spacing. A sphere is grown about the most gravitationally-bound dark matter particle of
each group until radii are found that enclose mean overdensities of $\Delta=94$,
$\Delta=200$, $\Delta=500$, $\Delta=1000$ and $\Delta=2500$, relative to the critical
density $\rho_\mathrm{cr,0}$. Any clusters which overlap with a more massive system within
these radii are discarded from the catalogues.

To check that dark matter structures are indeed undisturbed by the massless gas particles,
we have compared the positions of cluster centres in catalogues generated from our dark
matter-only simulation in the $L=62.5h^{-1}$ Mpc box and a non-radiative hydrodynamical
simulation in the same volume. At each overdensity $\Delta$, we find that the distance
between corresponding cluster centres in the two catalogues is less than the softening
length for over $99\%$ of our identified objects. We have also verified that this is the
case at high redshift $z\approx 3$.

\subsection{Implementing star formation and feedback from galaxies}

\subsubsection{Star formation}
\label{sec:starform}

Each model galaxy in the semi-analytic catalogues has an associated stellar mass $M_*$. To
compute the mass locked-up in stars formed over a redshift interval $\Delta
z=z_{n+1}-z_n$, we take the stellar mass of the galaxy at $z_n$ and subtract the sum of
the stellar masses of its progenitors at the previous output $z_{n+1}$:

\begin{equation}
\label{eq:Dmstar}
\Delta M_*=M_*(z_n)-\sum_\mathrm{prog.}M_*(z_{n+1}).
\end{equation} 

Once an output redshift is reached in a simulation, we convert the $\Delta N_{\rm star}=\Delta M_*/m_\mathrm{gas}$ gas particles nearest to each model galaxy into collisionless star particles, reflecting the increase in stellar mass $\Delta M_*$ of the galaxy since the last output. To ensure that $\Delta N_\mathrm{star}$ is an integer, we draw a random number $r$ uniformly from the unit interval and compare it with the fractional part of $\Delta N_\mathrm{star}$: if $r$ is less (greater) than the fractional part of $\Delta N_\mathrm{star}$, we round $\Delta N_\mathrm{star}$ up (down) to the nearest integer. Note that the star particles also have zero gravitational mass, so they do not influence the dark matter distribution.

\subsubsection{Type II supernova feedback}

The energy released by Type II SNe between two successive outputs depends on the amount of
mass that went into new stars during this time period. For each model galaxy, this is
simply $\Delta m_*=\Delta M_*/(1-R)$, where $\Delta M_*$ is obtained from equation
(\ref{eq:Dmstar}). Inserting $\Delta m_*$ into equation (\ref{eq:mej}) gives the mass of
ejected gas, corresponding to an energy input into the ICM of

\begin{equation}
\label{eq:Eej}
\Delta E_\mathrm{ejected}=\frac{1}{2}\Delta m_\mathrm{ejected}v_\mathrm{vir}^2
\end{equation}

\begin{figure}
\plotone{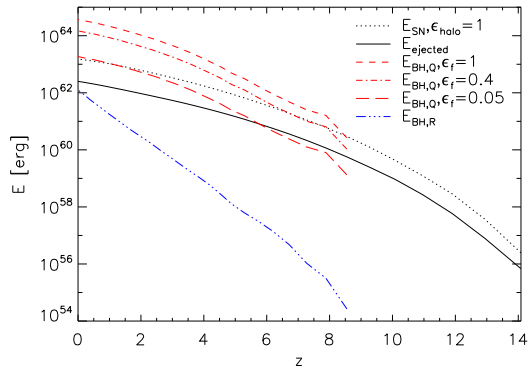}
\caption{Cumulative total amount of energy transferred to the ICM by different
  feedback sources as a function of redshift in a box of side length $L=62.5h^{-1}$ Mpc. The dotted line shows the total energy
  released by supernova explosions, while the solid line is the fraction of this
  energy available to heat intracluster gas. The contributions from quasar mode AGN
  feedback when the thermal coupling efficiency is $\epsilon_\mathrm{f}=1$, $\epsilon_{\rm
    f}=0.4$ and $\epsilon_\mathrm{f}=0.05$ are given by the dashed, dot-dashed and long dashed lines, respectively. The triple-dot-dashed line shows the mechanical heating associated with
  radio mode accretion. However, in the Munich L-Galaxies semi-analytic model, this energy
  only reduces the rate at which gas can cool out of the hot halo, rather than heating the
  ICM.}
\label{fig:Efeed}
\end{figure}

\n over the interval $\Delta z$. The solid line in Fig. \ref{fig:Efeed} shows the
L-Galaxies prediction for the cumulative total energy transferred to the ICM by Type II
SNe as a function of redshift. For comparison, the maximum possible total energy available
from SNe is also shown as the dotted line. This is computed at each redshift by
cumulatively summing $\Delta E_\mathrm{SN}$ for all galaxies, assuming an efficiency
$\epsilon_\mathrm{halo}=1$.

\placefigure{fig:Efeed}

At each output, we distribute the available energy $\Delta E_\mathrm{ejected}$ amongst the neighbouring gas particles of a galaxy using a simple heating model. The basis of our model is that all gas within a distance $r_{\rm vir}$ of a galaxy can be heated by feedback processes in the time $\Delta t$ between two outputs. We choose to inject energy in a distributed, rather than local, manner since, over a time $\Delta t$, heated gas will flow outwards, mixing with infalling cooler gas at larger radii. We have also experimented with alternative heating models where each galaxy heats a fixed number of neighbouring gas particles and found that our results are not significantly affected by the choice of heating model. In a forthcoming paper we intend to investigate different ways of injecting energy into the ICM in more detail.

We implement our heating model in GADGET-2 as
follows. For each galaxy, we find the number $N_\mathrm{ngb}$ of gas particles within a sphere
of radius $r_{\rm vir}$ centred on the galaxy. If no neighbours are found, the search
radius is increased until one gas particle is found. This is typically only necessary for low mass haloes. The energy released by SNe since the
last output is then used to raise the entropy of the neighbouring gas particles by a fixed
amount

\begin{equation}
\label{eq:DA}
\Delta A_i=\frac{(\gamma-1)\Delta E_\mathrm{ejected}}{m_\mathrm{gas}\sum_{j=1}^{N_{\rm
      ngb}}[\max{(f_\mathrm{b}\rho_\mathrm{vir},\rho_j)}]^{\gamma-1}},
\end{equation}

\n where $A=kT/(\mu m_\mathrm{p}\rho^{\gamma-1})$ is the definition of entropy employed by
GADGET-2, $m_\mathrm{p}$ is the mass of a proton and $\mu\approx 0.6$ is the mean
molecular weight for a fully-ionised gas of primordial composition. By giving each
particle a fixed entropy, rather than energy, boost, we ensure that denser particles close
to the galaxy are heated to a higher temperature than more distant, lower density
particles. The product of the cosmic baryon fraction $f_\mathrm{b}$ and the virial density
$\rho_\mathrm{vir}$ gives the mean overdensity of baryons within the virial radius. If no
neighbours are found within a distance $r_\mathrm{vir}$ of a galaxy and the search radius
has to be increased until one particle is found, the density of this particle may be less
than $f_\mathrm{b}\rho_{\rm vir}$. By using
$[\max{(f_\mathrm{b}\rho_\mathrm{vir},\rho_j)}]^{\gamma-1}$, rather than
$\rho_j^{\gamma-1}$, in the sum in equation (\ref{eq:DA}), we are assuming that the amount
of energy used to heat such particles is $\Delta
E_\mathrm{ejected}(\rho_i/f_\mathrm{b}\rho_{\rm vir})^{\gamma-1}<\Delta
E_\mathrm{ejected}$; the rest of the energy is taken to be used up as the gas does work
expanding adiabatically to a density $\rho_i<f_\mathrm{b}\rho_{\rm vir}$.

\subsubsection{Feedback from Active Galactic Nuclei}

A similar approach is used to calculate the energy released by AGN activity over a
redshift interval $\Delta z$. For each galaxy, the total change in mass of the central
black hole $\Delta M_\mathrm{BH}$ is given by equation (\ref{eq:Dmstar}), but with $M_*$
replaced by $M_\mathrm{BH}$. The fraction of this mass accreted via the radio mode is
approximately

\begin{equation}
\Delta M_\mathrm{BH,R}=\dot{M}_\mathrm{BH,R}\Delta t,
\end{equation}

\n where $\dot{M}_\mathrm{BH,R}$ is obtained by evaluating equation (\ref{eq:radio}) at
the current output $z_n$. It follows that the mass change due to merger-driven accretion
is then

\begin{equation}
\Delta M_\mathrm{BH,Q}=\Delta M_\mathrm{BH}-\Delta M_\mathrm{BH,R}.
\end{equation}

\n We ensure that the radiated luminosity

\begin{equation}
L_\mathrm{BH,Q}=\epsilon_\mathrm{r}\dot{M}_\mathrm{BH,Q}c^2
\end{equation}

\n does not exceed the Eddington luminosity

\begin{equation}
L_\mathrm{Edd}=1.3\times 10^{38}\left(\frac{M_\mathrm{BH}}{M_{\sun}}\right)\;{\rm
  erg\;s}^{-1}
\end{equation}

\n (e.g. \citealt{BEM82}) when averaged over the time $\Delta t$ between two successive
outputs. We then assume that some fraction of the energy radiated during quasar mode
accretion is coupled thermally to the ICM:

\begin{equation}
\label{eq:EBHQ}
\Delta E_\mathrm{BH,Q}=\epsilon_\mathrm{f}L_\mathrm{BH,Q}\Delta t,
\end{equation}

\n where the coupling efficiency $\epsilon_\mathrm{f}$ is a free parameter; we discuss
suitable values for this parameter in the following section. The dashed, dot-dashed and long dashed lines in Fig. \ref{fig:Efeed} show the cumulative total energy transferred to the ICM
by quasar mode AGN feedback for the cases $\epsilon_\mathrm{f}=1$,
$\epsilon_\mathrm{f}=0.4$ and $\epsilon_\mathrm{f}=0.05$, respectively. We inject the
energy $\Delta E_\mathrm{BH,Q}$ released by quasar mode accretion into gas particles
surrounding model galaxies using the same heating model as for supernova feedback.

The mechanical heating

\begin{equation}
\label{eq:EBHR}
\Delta E_\mathrm{BH,R}=\epsilon_\mathrm{r}\Delta M_\mathrm{BH,R}c^2
\end{equation}

\n associated with quiescent accretion simply reduces the rate at which gas in the hot
halo cools onto the disk. In other words, radio mode feedback does not explicitly heat
intracluster gas within the framework of L-Galaxies and is thus irrelevant for our hybrid
approach. For completeness, we show the cumulative total energy liberated by radio mode
accretion as the triple-dot-dashed line in Fig. \ref{fig:Efeed}.

\section{RESULTS AND DISCUSSION}
\label{sec:results}

\subsection{Model discrimination}

One of the most fundamental properties of galaxy groups and clusters is the X-ray
luminosity-temperature relation. Only recently have self-consistent hydrodynamical
simulations been able to successfully reproduce the slope and normalisation of this relation over a wide range of mass
scales. In this section we explore the individual effects of stellar and AGN feedback on
the \lxtx\ relation with a set of test simulations performed in the $L=62.5 h^{-1}$ Mpc
box using our hybrid technique.

\subsubsection{Star formation and supernova feedback}

\begin{figure}
\plotone{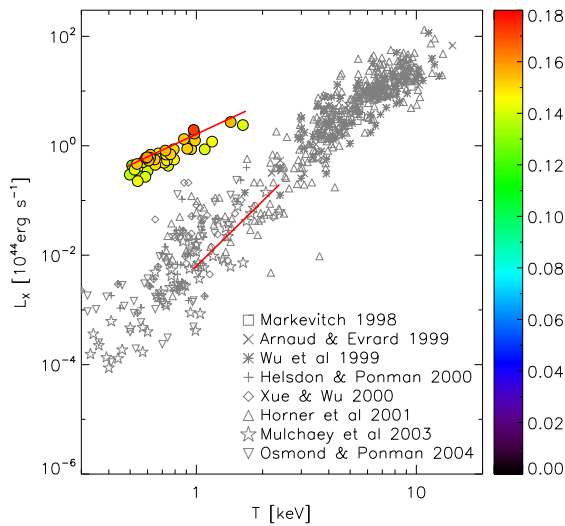}
\caption{Bolometric X-ray luminosity as a function of emission-weighted temperature for
  groups and clusters formed in a hybrid simulation with L-Galaxies star formation and
  Type II supernova feedback. X-ray properties are computed within $r_{500}$ and each
  object is shaded according to its hot gas fraction. For comparison, the upper and
  lower solid lines are the best fit relations from non-radiative and preheating
  simulations, respectively. Several datasets obtained from X-ray observations of groups
  and clusters are also shown.}
\label{fig:LTsfsn}
\end{figure}

As a starting point, we include only star formation and associated supernova feedback as
predicted by L-Galaxies. Fig. \ref{fig:LTsfsn} shows our simulated groups and clusters on
the \lxtx\ plane at $z=0$, shaded by their hot gas fraction. Bolometric X-ray
luminosities and emission-weighted temperatures were computed using the procedure outlined
by \citet{MTK02}. Since radiative cooling is neglected in our simulations, we do not
anticipate excess emission in cluster cores due to strong cooling flows and it is
therefore unnecessary to remove emission from central regions. All cluster properties are
calculated within $r_{500}$ since $\Delta=500$ is typically the smallest density contrast
accessible to observations. Furthermore, we only consider galaxy groups and clusters with
a mass $M_{500}\geq 1.3\times 10^{13}h^{-1}M_{\sun}$, corresponding to a total of about
$15,000$ particles within $r_{500}$. The most massive object formed has a mass
$M_{500}\approx 7\times 10^{13}h^{-1}M_{\sun}$.

\placefigure{fig:LTsfsn}

For comparative purposes, data from a number of X-ray observational studies is shown
\citep{MAR98,ARE99,WXF99,HEP00,XUW00,HOR01,MDM03,OSP04}. The uppermost solid line in
Fig. \ref{fig:LTsfsn} is the best-fit \lxtx\ relation obtained from a simulation that
includes gravitational heating only. We find that $\lx\propto T^{1.91\pm 0.37}$ in this
case, close to the $\lx\propto T^2$ scaling predicted by theoretical arguments. The
lower solid line shows the best-fit \lxtx\ relation for a simulation with uniform
preheating at high-redshift. The preheating model adopted imposes an entropy floor
$S_\mathrm{preheat}=200$ keV cm$^2$ at $z=4$. Preheating leads to a much steeper relation:
$\lx\propto T^{3.87\pm 0.87}$, consistent with observational data.

Observe that nearly all of our data points lie just below the best-fit
\lxtx\ relation for the gravitational heating simulation, with only a slight
hint of steepening. The hot gas fractions of our groups and clusters are at
least $65\%$ of the cosmic baryon fraction, with some having a hot gas fraction
very close to $f_{\rm b}$. Observational data suggests that only massive
clusters ($T\gtrsim 5$ keV) have such large gas fractions, with groups and poor
clusters typically having much smaller gas fractions
\citep{SPF03,VKF06,SVD09}. These results indicate that the entropy of the ICM
has been raised by stellar feedback, but nowhere near enough to explain observed
X-ray luminosities. 

By contrast, hydrodynamical simulations with radiative
cooling, star formation and supernova feedback tend to produce a closer match to
the observed \lxtx\ relation (e.g. \citealt{BMS04,PSS08}). There are two reasons
for this. Firstly, the amount of baryons that cool and form stars is typically
much greater in hydrodynamical simulations, with $30-50\%$ of the baryons within
the virial radius locked-up in stars (e.g. \citealt{BPB01,DKW02,TBS03}). In our
hybrid simulation, the average stellar fraction within the virial radius is approximately
$9\%$, in agreement with observational data \citep{BPB01,LMS03,BMBE08}. This is,
of course, to be expected since star formation in our simulation is driven by a
SAM which has been tuned to reproduce the cosmic star formation
history. Secondly, popular stellar feedback schemes typically assume that each
supernova event transfers more energy to the surrounding gas. For example, in
the models of \cite{SPH03} and \citet{KTJ04}, the amount of energy returned to
the ICM per solar mass of stars formed is at least $4\times 10^{48}$ erg
$M_{\sun}^{-1}$. Within the framework of L-Galaxies, it follows from equation
(\ref{eq:mej}) that this is only possible if no energy is used to reheat cold
disk gas ($\Delta E_\mathrm{hot}=0$) and the efficiency parameter
$\epsilon_\mathrm{halo}=1$.

\begin{figure}
\plotone{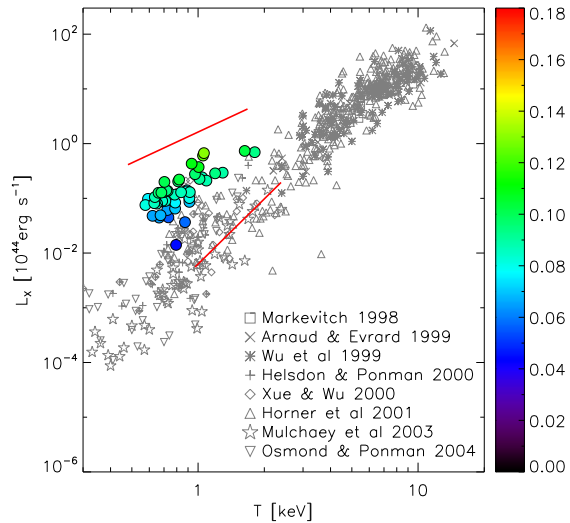}
\caption{Same as Fig. \ref{fig:LTsfsn}, but with an enhanced stellar feedback scheme in
  which the L-Galaxies star formation efficiency is $5$ times greater and the amount of
  energy transferred to the ICM by Type II SNe has been set to its maximum possible
  value.}
\label{fig:LTsf5sn5}
\end{figure}

To test that our method yields comparable results to existing hydrodynamical simulations
when the star formation efficiency is increased, we have performed a simulation where the
mass of newly formed stars $\Delta m_*$ is simply multiplied by $5$. We
also assume the energy imparted to the ICM by SNe is larger: $\Delta E_{\rm
  ejected}=\Delta E_\mathrm{SN}$, with an efficiency $\epsilon_\mathrm{halo}=1$. Note that
setting $\Delta E_\mathrm{hot}=0$ means all star-forming galaxies can eject gas from their
halo, rather than just those with a virial velocity $v_\mathrm{vir}\lesssim 200$ km
s$^{-1}$. Fig. \ref{fig:LTsf5sn5} shows the resulting \lxtx\ relation. It is evident that
cluster X-ray luminosities and hot gas fractions are significantly reduced relative to
those predicted by the standard L-Galaxies stellar feedback model. Our results are in
broad agreement with those obtained from simulations incorporating star formation and
supernova feedback (e.g. \citealt{BMS04,PSS08}). However, the normalisation of the
\lxtx\ relation is still too high and the slope is too shallow relative to the observed
relation. The stellar fraction within the virial radius has now increased to $39\%$ on average,
conflicting with observations but again similar to results from direct simulations
(e.g. \citealt{PSS08}).

\placefigure{fig:LTsf5sn5}

The heating of intracluster gas by stellar feedback from model galaxies is clearly
insufficient to reproduce the \lxtx\ relation, particularly on group scales. This remains
the case even when the star formation efficiency is unrealistically high. Even if we were
to allow for the energy released by Type Ia SNe, this is unlikely to bring the
\lxtx\ relation shown in Fig. \ref{fig:LTsfsn} in line with observations since the
energetics of Type Ia and Type II SNe are thought to be roughly similar. However, Type Ia
SNe are crucially important for chemically enriching the ICM at low-redshift
(e.g. \citealt{MDP05}); an issue we shall address elsewhere. We now investigate whether
additional energy input from AGNs can resolve this problem, starting with feedback from
quasar-induced outflows.

\subsubsection{Including feedback from Active Galactic Nuclei}

Cosmological simulations including prescriptions for black hole growth and
associated quasar activity have been performed by \citet{DCS08}. They assumed
$5\%$ of the energy radiated during quasar mode accretion is coupled thermally
and isotropically to the surrounding medium. This choice ensures that the
normalisation of the local black hole-bulge mass relation agrees with
observations \citep{DSH05}. In their model, the radiated luminosity is related to
the total black hole accretion rate $\dot{M}_\mathrm{BH}$, whereas we assume it
is governed by the merger-driven accretion rate $\dot{M}_\mathrm{BH,Q}$. However, the
growth of black holes in L-Galaxies is dominated by quasar mode accretion, so
that $\Delta M_\mathrm{BH}\approx\Delta M_\mathrm{BH,Q}$ anyway. Based on this
argument, it seems sensible to begin by setting $\epsilon_\mathrm{f}=0.05$ in
equation (\ref{eq:EBHQ}).

\begin{figure}
\plotone{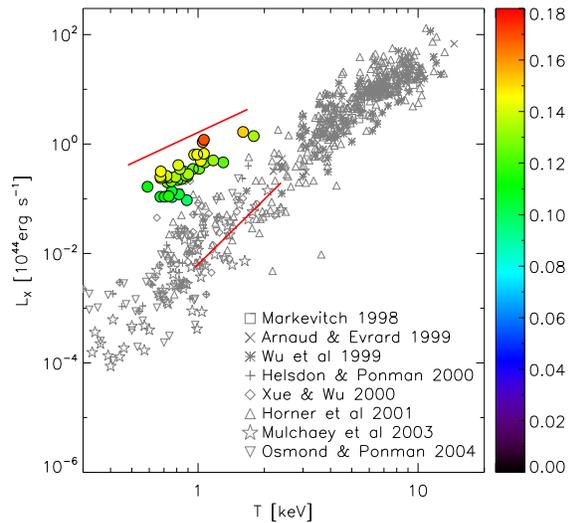}
\caption{Same as Fig. \ref{fig:LTsfsn}, but for a hybrid simulation with L-Galaxies
  stellar feedback and a simple model for quasar mode AGN feedback. The energy released by
  merger-driven accretion onto the central black hole is assumed to be thermally coupled
  to the intracluster gas with an efficiency $\epsilon_\mathrm{f}=0.05$.}
\label{fig:LTsfsnquas0.05}
\end{figure}

Fig. \ref{fig:LTsfsnquas0.05} shows the \lxtx\ relation obtained from a hybrid
simulation including star formation, supernova and quasar mode AGN feedback with
a coupling efficiency $\epsilon_\mathrm{f}=0.05$. Again, cluster X-ray
luminosities and hot gas fractions are considerably over-estimated. \citet{DCS08}
did not use their model to investigate the \lxtx\ relation of groups and
clusters. However, since the global black hole accretion rate histories in
L-Galaxies and the simulations of \cite{DCS08} are similar, we would expect them
to obtain a similar result.

\placefigure{fig:LTsfsnquas0.05}

\begin{figure}
\plotone{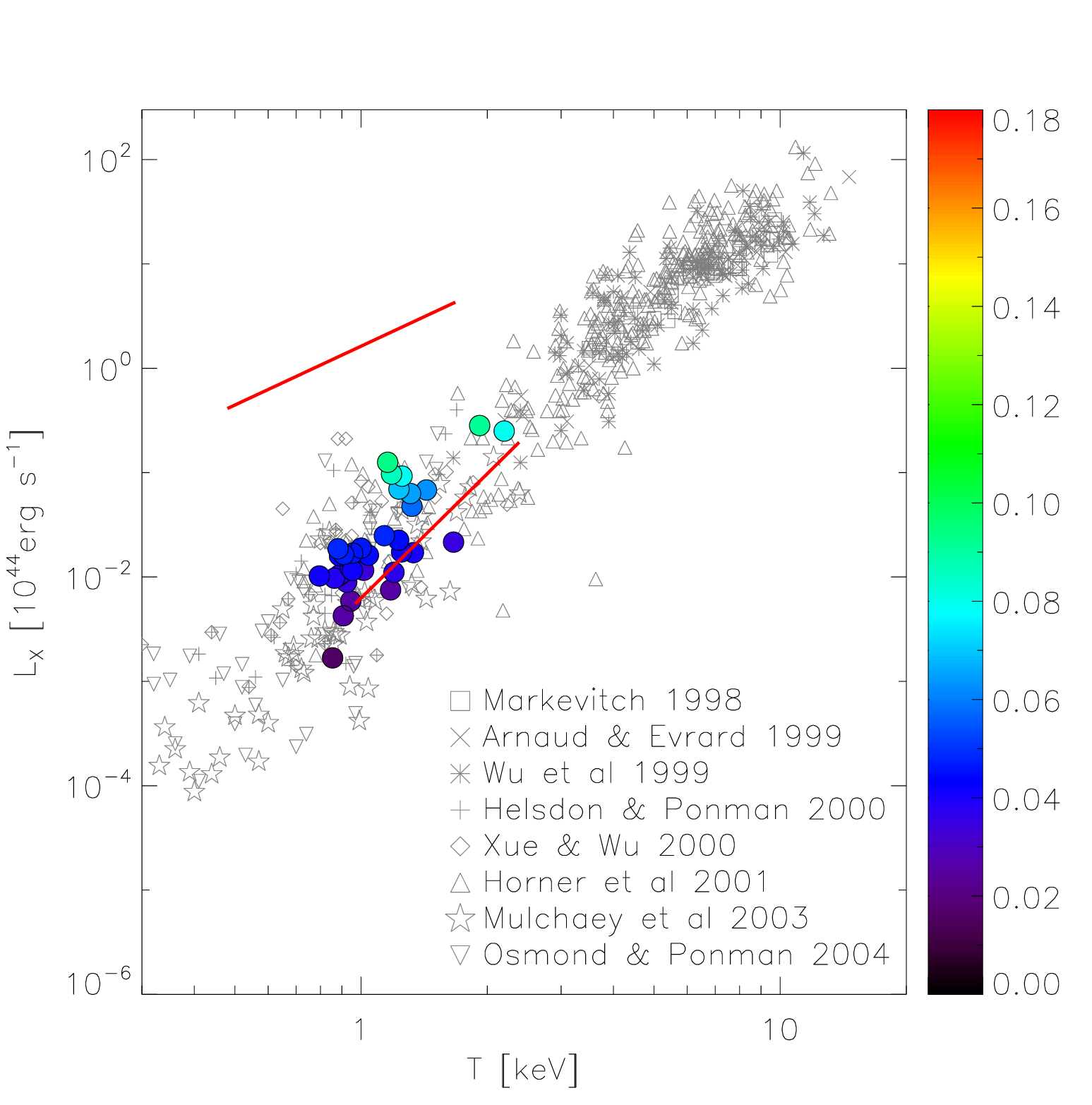}
\caption{Same as Fig. \ref{fig:LTsfsnquas0.05}, except the thermal coupling efficiency for
  quasar mode AGN feedback has been increased to $\epsilon_\mathrm{f}=0.4$.}
\label{fig:LTsfsnquas0.4}
\end{figure}

In order to recover the observed steepening of the \lxtx\ relation on group scales, we
find that the quasar mode coupling efficiency has to be increased to
$\epsilon_\mathrm{f}=0.4$. The \lxtx\ relation in this case is shown in
Fig. \ref{fig:LTsfsnquas0.4}. Our simulated clusters now provide a much better match to
the observational data. In addition, the gas fractions of our clusters are in
broad agreement with observations \citep{SPF03,SVD09}.

\placefigure{fig:LTsfsnquas0.4}

Artificially boosting the feedback from merger-driven accretion can be viewed as a crude
attempt at including radio mode feedback since, at each redshift, over $75\%$ of model
galaxies with a central black hole accrete via both modes. Can we recover the correct
\lxtx\ relation by explicitly including the radio mode contribution, rather than simply
increasing $\epsilon_\mathrm{f}$? Recall that the L-Galaxies implementation of radio mode
feedback only reduces the rate at which gas can cool out of the quasi-static hot halo,
rather than directly heating the ICM. Even if we temporarily ignore this feature of the
model and inject all of the energy $\Delta E_\mathrm{BH,R}$ (see equation \ref{eq:EBHR})
released by quiescent accretion into the gas component of a simulation incorporating
stellar and quasar mode feedback with $\epsilon_\mathrm{f}=0.05$, we find that it has
little effect on the \lxtx\ relation shown in Fig. \ref{fig:LTsfsnquas0.05}. This is
because the mechanical heating associated with radio mode accretion is negligible at high
redshift, only becoming comparable to the energy released by SNe at $z=0$; see
Fig. \ref{fig:Efeed}. However, \citet{PSS08} have successfully reproduced the observed mean \lxtx\ relation by incorporating a model for AGN-driven bubble heating into the quasar
mode feedback scheme of \citet{DCS08}. This highlights the need for an alternative model
of AGN feedback in which the radio mode does more than just offset cooling from the hot
halo.

In recent work, \citet{BMB08} have used the Durham SAM GALFORM to investigate the
properties of the ICM, particularly the \lxtx\ relation. They demonstrated that the
version of the model developed by \citet{BBM06} over-predicts X-ray luminosities on group
scales, leading to a shallow \lxtx\ relation similar to the one we obtained from our
simulation with L-Galaxies stellar feedback (Fig. \ref{fig:LTsfsn}). In this model, the
energy released by quiescent radio mode accretion simply prevents any significant amount
of gas cooling in massive haloes, as in L-Galaxies. To try to explain the X-ray properties
of galaxy groups and clusters whilst simultaneously accounting for the observed properties
of galaxies, \citet{BMB08} proposed a modification of GALFORM which allows for heat input
into the ICM from radio mode feedback. This additional heating acts to expel gas from
the X-ray emitting central regions of haloes, reducing the gas density and thus
luminosity. Lower-mass systems are affected more than massive ones because their cooling
time is shorter, meaning they initially supply more material to the central black hole,
resulting in a larger amount of feedback per unit mass of gas. The \lxtx\ relation then
becomes steeper as desired on group scales. With this modification, the GALFORM model reproduces the observed mean \lxtx\ relation and the substantial scatter about this relation at low temperatures ($T\lesssim 3$ keV), as well as halo gas fractions \citep{BMB08}.

The basis of the \citet{BMB08} AGN feedback model is to compute the heating power $L_{\rm
  heat}$ as a function of the cooling rate. More specifically,

\begin{equation}
\label{eq:Lheat}
L_\mathrm{heat}=\eta_\mathrm{SMBH}\epsilon_\mathrm{r}\dot{M}_\mathrm{cool}c^2,
\end{equation}

\n subject to the constraint

\begin{equation}
\label{eq:Eddlim}
L_\mathrm{heat}\leq\epsilon_\mathrm{SMBH}L_\mathrm{Edd}.
\end{equation}

\n Here $\dot{M}_\mathrm{cool}$ is the cooling rate of the halo in the absence of radio
mode feedback, related to the black hole accretion rate via $\dot{M}_\mathrm{BH}=\eta_{\rm
  SMBH}\dot{M}_\mathrm{cool}$. The parameter $\eta_\mathrm{SMBH}$ controls the efficiency
with which cooling material can be accreted by the black hole and is set to $\eta_{\rm
  SMBH}=0.01$ by \citet{BMB08}. The limiting criterion (\ref{eq:Eddlim}) relates to the
structure of the accretion disk itself. Effective radio mode feedback requires efficient
jet production, typically thought to be associated with geometrically thick, advection
dominated disks (e.g. \citealt{RBB82,MEI01,CSS05}). If the accretion rate is too high,
models suggest that the vertical height of the disk will collapse, leading to a drop in
jet production efficiency. Much more of the energy released by accretion is then radiated
away and is not available for radio mode feedback. Based on theoretical work
(e.g. \citealt{EMN97}), \cite{BMB08} assume this change in structure of the accretion disk
occurs once the accretion rate reaches
$\dot{M}_\mathrm{BH}=\epsilon_\mathrm{SMBH}\dot{M}_{\rm Edd}$, where
$\epsilon_\mathrm{SMBH}$ is referred to as the disk structure parameter and is related to
the disk viscosity parameter $\alpha$ by $\epsilon_{\rm SMBH}\propto\alpha^2$. The disk
structure parameter is assigned a value $\epsilon_{\rm SMBH}=0.02$, in accordance with
plausible accretion disk viscosities (e.g. \citealt{MCG04,HKD04,HAK06}). Note that the AGN
feedback model suggested by \citet{SSD07} also assumes that radio mode feedback is only
effective for accretion rates below some fraction of the Eddington rate.

The heating energy available from the AGN is compared with the energy lost radiatively by gas
cooling from the halo. If the feedback energy is greater than the radiated energy, the
excess energy is used to eject gas from the halo. The energy transferred to the ICM over a
time period $\Delta t$ is then

\begin{equation}
\label{eq:AGNEej}
\Delta E_\mathrm{ejected}=\Delta E_\mathrm{heat}-\Delta E_\mathrm{cool},
\end{equation}

\n where $\Delta E_\mathrm{heat}=L_\mathrm{heat}\Delta t$ and

\begin{equation}
\label{eq:Ecool}
\Delta E_\mathrm{cool}=\frac{1}{2}\Delta M_\mathrm{cool}v_\mathrm{vir}^2.
\end{equation}

We have implemented the \citet{BMB08} prescription for AGN feedback in our hybrid approach
using the following procedure. For each galaxy in the catalogue, we take the total change
in black hole mass between two successive outputs, $\Delta M_\mathrm{BH}$, and compute
$\Delta M_\mathrm{cool}=\Delta M_\mathrm{BH}/\eta_\mathrm{SMBH}$. Inserting this into
equations (\ref{eq:Lheat}) and (\ref{eq:Ecool}) gives $\Delta E_\mathrm{heat}$ and $\Delta
E_{\rm cool}$, respectively. The heat input into the ICM from each model galaxy then
follows from equation (\ref{eq:AGNEej}). This energy is injected into the gas component of
our simulation using the same heating model as described previously. 

To be fully consistent, we should take into account the effect of AGN heating on hot halo gas within L-Galaxies itself. As in \citet{BMB08}, this would almost certainly lead to changes in some of the other model parameters. However, if we were to adjust the appropriate parameters so as to regain a realistic galaxy population, then we would require galaxy properties (particularly their black hole masses) to be similar to those in our existing galaxy catalogue. The amount of heating energy available from AGNs would then also be (approximately) the same since it is driven purely by the black hole growth rate in the \citet{BMB08} model.

\begin{figure}
\plotone{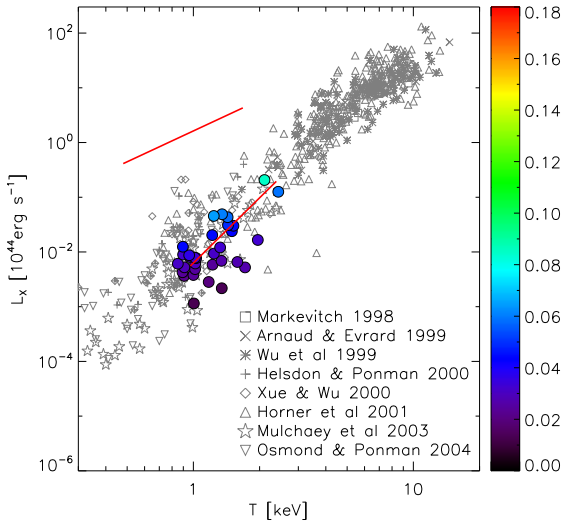}
\caption{Same as Fig. \ref{fig:LTsfsn}, but for a hybrid simulation with L-Galaxies
  stellar feedback and the \citet{BMB08} AGN feedback prescription employed in the latest
  version of GALFORM.}
\label{fig:LTsfsnagn}
\end{figure}

Fig. \ref{fig:LTsfsnagn} shows the \lxtx\ relation obtained from a simulation with
L-Galaxies stellar feedback and the AGN feedback model of \citet{BMB08}. Our results
clearly provide an excellent match to the observational data. In addition, the hot gas and
stellar fractions of our groups and clusters agree with observed values. This success
indicates that the L-Galaxies implementation of AGN feedback must indeed be revised
if the model is to explain the observed properties of both the galaxy distribution and the
ICM.

\placefigure{fig:LTsfsnagn}

\subsection{Best-fit model}

Following the analysis presented in the preceding section, we have established a model
that is capable of reproducing the observed \lxtx\ relation of groups and poor
clusters. This best-fit model comprises of the stellar feedback scheme employed in
L-Galaxies, combined with the AGN feedback prescription of \cite{BMB08}. To investigate
whether this model can also correctly explain the properties of richer clusters, we have
performed a hybrid simulation in a larger $L=125h^{-1}$ Mpc volume.

\subsubsection{The X-ray luminosity-temperature relation}

\begin{figure}
\plotone{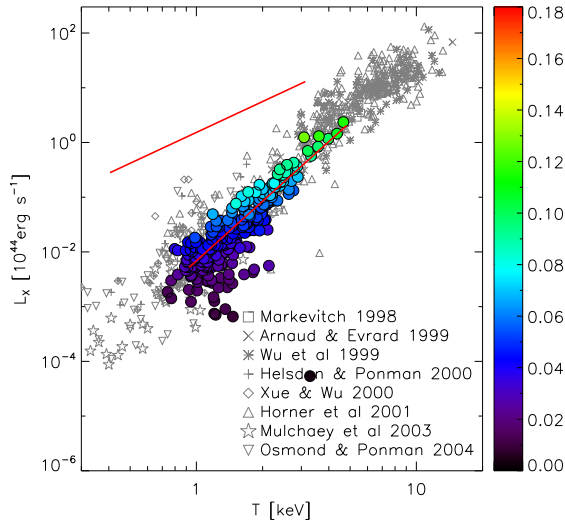}
\caption{Bolometric X-ray luminosity as a function of emission-weighted temperature for a
  hybrid simulation with our best-fit feedback model. This model consists of stellar
  feedback from L-Galaxies combined with the \citet{BMB08} AGN feedback prescription. See
  the caption of Fig. \ref{fig:LTsfsn} for a description of the shading and best-fit
  lines.}
\label{fig:LTsfsnagn125}
\end{figure}

The $z=0$ \lxtx\ relation obtained from our simulation is shown in
Fig. \ref{fig:LTsfsnagn125}. Groups and clusters are shaded by their hot gas
fraction and we only plot clusters with a mass $M_{500}\geq 1.3\times
10^{13}h^{-1}M_{\sun}$. The mass of the largest cluster formed is now $M_{500}\approx
2.3\times 10^{14}h^{-1}M_{\sun}$. For comparison, we show the best-fit \lxtx\ relations from
a non-radiative simulation (upper solid line) and a simulation with the same
preheating model as described previously (lower solid line). These relations are of
the form $\lx\propto T^{1.88\pm 0.16}$ and $\lx\propto T^{3.62\pm 0.33}$,
respectively. As before, our hybrid simulation yields an \lxtx\ relation with a slope and normalisation that is generally consistent with observations at all mass
scales. A few objects appear to have a higher temperature than expected, given their X-ray
luminosity. This is particularly true for the $T\approx 3.3$ keV cluster with a luminosity
$\lx\approx 5.4\times 10^{39}$ erg s$^{-1}$. The reason for this is that the gas in this
object has recently been raised to a high temperature by a large energy injection from the
central AGN, causing it to flow outwards. As this gas is replaced by infalling cooler gas,
the system will stabilise and shift back towards the main relation. Our results exhibit a
variation in scatter along the \lxtx\ relation that is similar to observational data. In
particular, for temperatures $T\lesssim 3$ keV, our data points fan out to populate a
triangular region of the \lxtx\ plane in the same way. This scatter about the
low-temperature end of the \lxtx\ relation is attributable to the variety of merger
histories of groups. Note that there seems to be more scatter towards the lower-luminosity edge of the observed relation than the upper-luminosity edge. This is because we cannot produce systems with a highly X-ray luminous CC in our simulations since we have chosen to neglect cooling processes (see Section \ref{sec:profiles}).


\placefigure{fig:LTsfsnagn125}

\subsubsection{Halo gas fractions}

\begin{figure}
\plotone{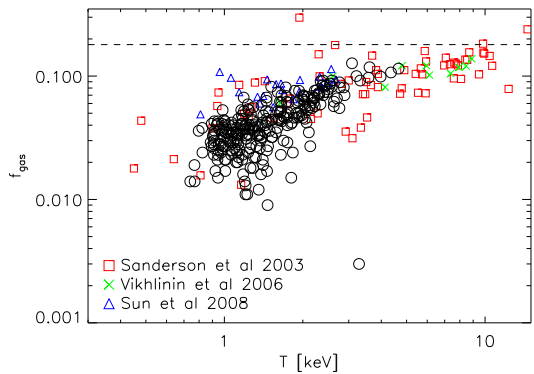}
\caption{Halo hot gas fraction within $r_{500}$ as a function of emission-weighted
  temperature for a hybrid simulation with our best-fit feedback model. Simulated objects
  are shown by the open circles. A variety of data from X-ray observations of groups
  and clusters is shown for comparison. The horizontal dashed line is the mean
  cosmic baryon fraction, $f_\mathrm{b}$, in the cosmological model we have adopted.}
\label{fig:gasfrac}
\end{figure}

It is evident from Fig. \ref{fig:LTsfsnagn125} that, as we move along the \lxtx\ relation
from group to cluster scales, the hot gas fraction of our simulated objects increases,
reaching approximately $70\%$ of the mean cosmic baryon fraction in massive clusters. To
illustrate this behaviour more clearly, we explicitly plot the hot gas fraction
$f_\mathrm{gas}$ as a function of emission-weighted temperature (both within $r_{500}$) in
Fig. \ref{fig:gasfrac}. For comparative purposes, we show constraints on halo gas
fractions obtained from X-ray observations \citep{VKF06,SVD09}. Also plotted are gas
fractions within $r_{500}$ computed from the gas density and temperature profile
parameters given in \citet{SPF03}. 

\placefigure{fig:gasfrac}

The gas fractions of our groups and clusters are in
broad agreement with the observational data, exhibiting a comparable amount of scatter. In
both our results and the data we see a rapid decline in gas fraction at lower
temperatures. This is because AGN feedback is effective at driving gas from the central
regions of low-mass systems. On the other hand, the potential wells of massive clusters
are too deep for AGN heating to efficiently remove gas from them, so they retain a much
larger fraction of their hot gas. Finally, we note that the $T\approx 3.3$ keV cluster
with the smallest gas fraction ($0.3\%$) corresponds to the system in
Fig. \ref{fig:LTsfsnagn125} with $\lx\approx 5.4\times 10^{39}$ erg s$^{-1}$ discussed
before. This supports our argument that this object has recently experienced an intense
burst of AGN activity which has driven a substantial amount of gas beyond $r_{500}$.

\subsubsection{Entropy and temperature profiles}
\label{sec:profiles}

\begin{figure}
\plotone{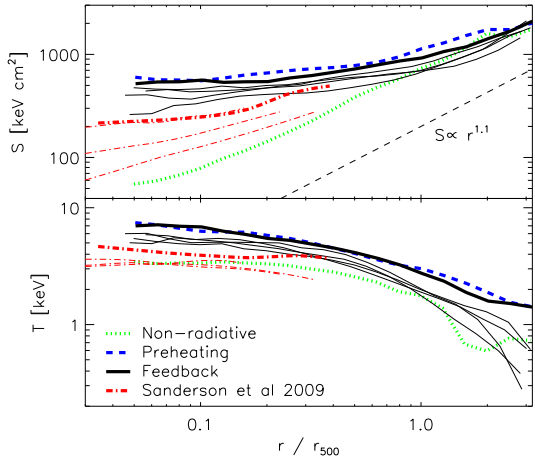}
\caption{Radial profiles of entropy (upper panel) and emission-weighted temperature (lower panel) obtained from a hybrid simulation with our best-fit feedback model. The solid lines are profiles for a sample of five clusters with a mass $1.1 \times 10^{14}h^{-1}M_{\sun}\leq M_{500}\leq 1.7\times 10^{14}h^{-1}M_{\sun}$, with the profiles of the most massive cluster in this sample highlighted by thick solid lines. For comparative purposes, the thick dotted and dashed lines are the profiles of the corresponding object formed in a non-radiative and a preheating simulation, respectively. In addition, observed profiles for four non-cool core clusters from the sample presented by \citet{SOP09} are shown as thin dot-dashed lines. These objects are of similar mass to our simulated clusters: $1.1 \times 10^{14}h^{-1}M_{\sun}\leq M_{500}\leq 1.6\times 10^{14}h^{-1}M_{\sun}$, and the thick dot-dashed lines highlight the profiles of the most massive object. Theoretical arguments predict that entropy scales as $S\propto r^{1.1}$ outside of cluster cores, shown by the thin dashed line in the upper panel.}
\label{fig:profs}
\end{figure}

Further information about non-gravitational heating processes operating in clusters can be gleaned
by inspecting radial entropy and temperature profiles of the ICM. In the upper (lower) panel of
Fig. \ref{fig:profs}, the thin solid lines are entropy (emission-weighted temperature) profiles for a sample of five of our most massive clusters, with masses in the range $1.1\times 10^{14}h^{-1}M_{\sun}\leq M_{500}\leq 1.7\times 10^{14}h^{-1}M_{\sun}$ at $z=0$. Thick solid lines highlight the profiles of the most massive cluster in this sample (the second most massive object in the simulation volume). For reference, the thick dotted and dashed lines are the
profiles of the corresponding object formed in the non-radiative and preheating
simulations, respectively. Profiles are only plotted for radii greater than the
gravitational softening length. In addition, observed profiles for four NCC clusters from the sample presented by \citet{SOP09} are shown as thin dot-dashed lines. These objects are of similar mass to our simulated clusters: $1.1\times 10^{14}h^{-1}M_{\sun}\leq M_{500}\leq 1.6\times 10^{14}h^{-1}M_{\sun}$, with the profiles of the most massive object highlighted by thick dot-dashed lines. We only compare our profiles with those of NCC clusters since our simulations are non-radiative and thus systems with a CC do not form. 

\placefigure{fig:profs}

Theoretical models of shock heating during spherical collapse predict that entropy scales
with radius as $S\propto r^{1.1}$ (e.g. \citealt{TON01}). This scaling behaviour is indeed
observed in cluster outskirts, but entropy profiles are typically seen to become flatter in central
regions $r\lesssim 0.2 r_{200}$ (e.g. \citealt{PSF03,PAP06}). However, the precise radius at which this flattening occurs varies considerably, depending on such factors as the temperature (mass) of the system and whether it has a CC or a NCC. In particular, hotter, more massive objects have a higher mean core entropy (e.g. \citealt{CDV09}), and the profiles of NCC clusters flatten off at significantly larger radii than those of CC clusters  (e.g. \citealt{SOP09}).

The power-law $S\propto r^{1.1}$ is illustrated by the thin dashed line in the upper panel of
Fig. \ref{fig:profs} (the normalisation is arbitrary). It is evident that the entropy
profiles of our clusters depart from this scaling at somewhat larger radii than
observed: $r\approx r_{500}$, flattening as we move in towards the core and
over-estimating the central entropy. Likewise, the temperature profiles shown
in the lower panel of Fig. \ref{fig:profs} reveal that the core temperature is higher than
would be expected for objects of this mass. This is similar to the behaviour
predicted by the preheating model, although our feedback model yields an entropy
profile that provides a slightly better match to the observational data.

Based on the preceding discussion it is natural to ask how we can recover the
observed \lxtx\ relation and halo gas fractions within $r_{500}$ if the gas
entropy is over-estimated in cluster cores.  The answer lies in the fact that
the full \citet{SOP09} sample contains clusters with a variety of entropy profiles.  If
we consider the emission per unit logarithmic radius, ${\rm d}L_{\rm
  X}/{\rm d}\log_{10}r$, the most centrally-concentrated CC clusters have their
peak emission within $0.2r_{500}$, whereas in some NCC clusters the peak is beyond $0.7r_{500}$.
Our most-massive simulated clusters resemble these latter objects, and hence lie
at the lower-luminosity edge of the \lxtx\ relation (for low-mass systems, the
observations do not extend far enough in radius to make a meaningful comparison).


We note that profiles can only be reliably measured for groups and poor clusters that are very X-ray bright. However, optically-selected samples of groups have revealed systems with little or no detectable X-ray emission that may be the group analogue of NCC clusters (e.g. \citealt{RPM06}). It may not be possible to extract profiles for such objects, but since they are X-ray under-luminous we would expect them to have a large entropy core, possibly similar to that seen in the profiles of our simulated groups.

However, in order to reproduce the observed profiles of the majority of X-ray bright groups and poor clusters, the entropy and temperature of the gas in core regions must be lowered. This could potentially be achieved by incorporating cooling processes in our simulations. The inclusion of radiative cooling would only strengthen our conclusion that large amounts of energy must be injected into the ICM by AGNs to recover the observed \lxtx\ relation and halo gas fractions.

\section{RESOLUTION TEST}
\label{sec:restest}

In all of the hydrodynamical simulations presented in this paper we have chosen a lower
resolution for the gas component than for the dark matter. To check that this approach
yields robust results, we have performed a suite of six hybrid simulations in the
$L=62.5h^{-1}$ Mpc box with a fixed number $N_\mathrm{DM}=270^3$ of dark matter particles,
but different numbers of gas particles. Two of these simulations have
$N_\mathrm{gas}<100^3$, while for the other three $N_\mathrm{gas}>100^3$ (recall that all
our simulations in this volume have $N_\mathrm{gas}=100^3$). All other simulation
parameters are unchanged between the runs. The feedback scheme we adopt is our best-fit
model since this is the only one that we have used in the larger $L=125h^{-1}$ Mpc box and
it is this model we intend to apply in the future.

\begin{figure}
\plotone{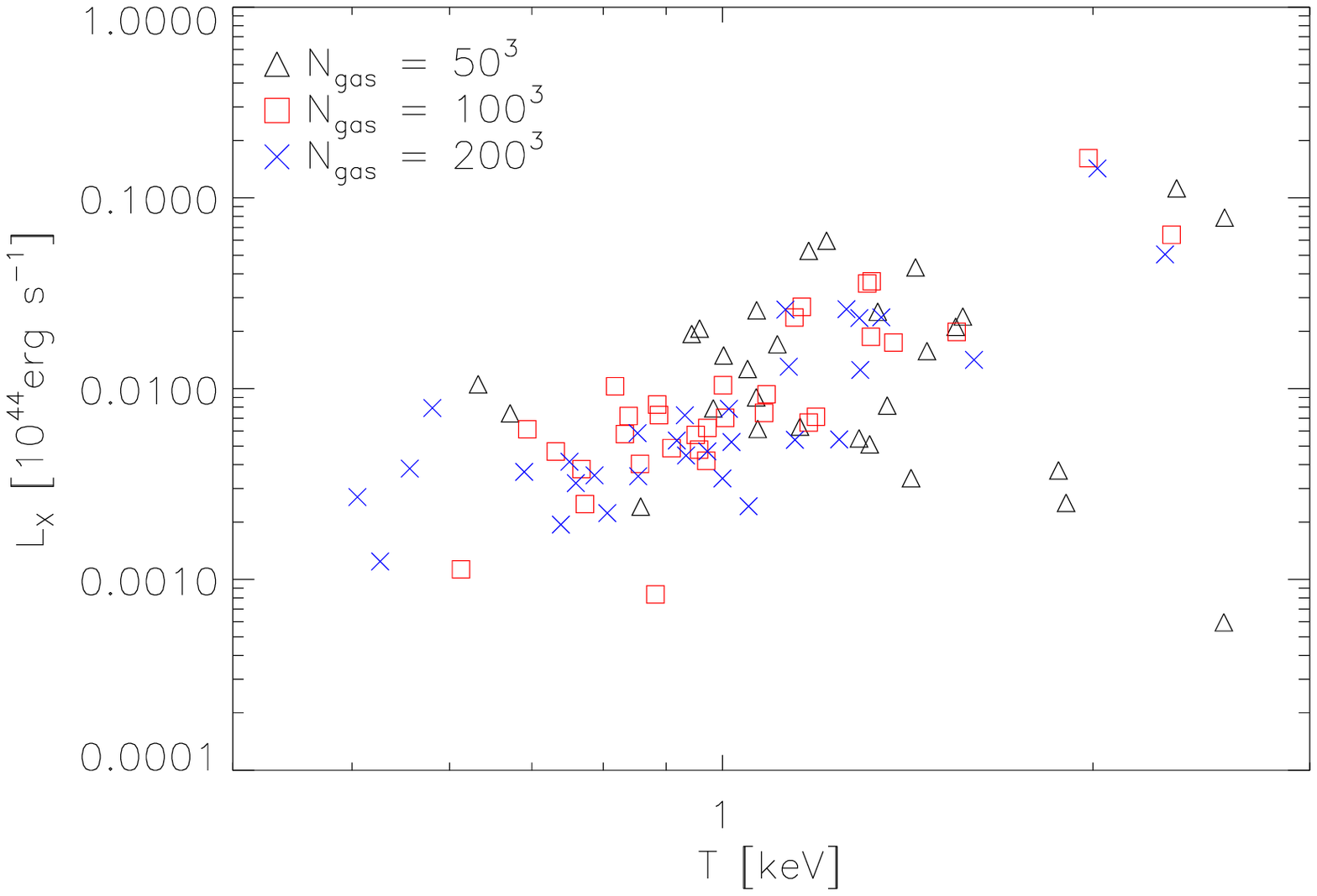}
\caption{Bolometric X-ray luminosity as a function of emission-weighted temperature for
  three hybrid simulations with our best-fit feedback model. To demonstrate the effect of
  resolution, the simulations are identical in all aspects except for the number of gas
  particles used, as detailed in the legend.}
\label{fig:LTrestest}
\end{figure}

To demonstrate that $N_\mathrm{gas}=100^3$ is a sufficient number of gas particles for a
converged estimate of the \lxtx\ relation, in Fig. \ref{fig:LTrestest} we compare the
relations obtained from three simulations with $N_\mathrm{gas}=50^3$,
$N_\mathrm{gas}=100^3$ and $N_\mathrm{gas}=200^3$, respectively. The corresponding (true)
masses of the gas particles are then $m_\mathrm{gas}=2.44\times10^{10}h^{-1}M_{\sun}$, $m_\mathrm{gas}=3.05\times 10^9h^{-1}M_{\sun}$, and $m_\mathrm{gas}=3.81\times
10^8h^{-1}M_{\sun}$. X-ray luminosities and emission-weighted temperatures are computed
within $r_{500}$ and only objects with a mass $M_\mathrm{500}\geq 1.3\times
10^{13}h^{-1}M_{\sun}$ are plotted. Focusing on the two most massive clusters, we see that
they have very similar temperatures and luminosities in the intermediate and
high-resolution runs. We find that this remains the case if we go to even higher
resolution. However, they shift significantly in the low-resolution simulation. This
signifies numerical convergence at the intermediate resolution for systems with $T\gtrsim
2$ keV, which is important since we plan to use our technique to preferentially study rich
clusters in subsequent work.

\placefigure{fig:LTrestest}

As we move to lower temperatures, the amount of scatter between the three different
resolution simulations increases. In the low resolution run, the data points are less
tightly grouped and seem to lie on a mean relation that has a shallower slope than in the
other two simulations. In addition, three objects have anomalously high temperatures given
their X-ray luminosity. This is because there is a limited number of gas particles
available for heating in these low-mass systems at this resolution and a large amount of
energy has recently been injected into a small number of gas particles. On the other hand,
the groups and clusters formed in the intermediate and high-resolution runs lie on a
similar mean relation, although the data points do not line-up perfectly. This indicates
that we have not attained numerical convergence at the lowest resolution. Note that the
\lxtx\ relation in the intermediate and high-resolution runs appears to flatten slightly
at the low temperature end in the same way as observational data.

It is not surprising that we do not achieve exact convergence for two reasons. Firstly, in
situations where only one gas particle is heated, it is evident from equation
(\ref{eq:DA}) that the entropy boost given to the particle is inversely proportional to
its (true) mass: $\Delta A_i\propto 1/m_\mathrm{gas}$, if its density is less than $f_{\rm
  b}\rho_\mathrm{vir}$. Consequently, as the resolution is increased, the particle will
receive a larger entropy injection. Secondly, there is a stochastic element to our star
formation scheme. By performing another intermediate-resolution simulation with a
different random seed, we found that this randomness induces considerable scatter in the
luminosities and temperatures of the lowest-mass systems, but has little effect in more
massive objects. Based on these considerations, we suggest that the X-ray properties
computed for groups with $T\lesssim 1$ keV are unreliable since these systems contain
relatively few gas particles. However, we feel that the level of convergence between the
intermediate and high-resolution runs for objects with $T\gtrsim 1$ keV is sufficient to
give a good statistical representation of the net effect of feedback on the $L_{\rm
  X}$-$T$ relation.

\section{SUMMARY AND CONCLUSIONS}
\label{sec:conc}

In this paper we set out to extend the predictive power of current semi-analytic models (SAMs) of galaxy
formation by investigating the effect of energy feedback from model galaxies on the
properties of intracluster gas. To achieve this objective we have employed a novel hybrid
technique in which a SAM is coupled to non-radiative hydrodynamical simulations, thus guaranteeing that the source of feedback in our simulations is a realistic galaxy population. This is the first time such an approach has been adopted and is complementary to existing theoretical studies of galaxy groups and clusters based on self-consistent hydrodynamical simulations.

The main result to emerge from our work is that a large energy input from AGNs (on average, $35\%$ of the available rest mass energy $\epsilon_{\rm r}M_{\rm BH}c^2$) is required over the entire formation history of haloes in order to reproduce the observed \lxtx\ relation and halo gas fractions. This supports the conclusion of \citet{BMB08} derived using purely semi-analytic reasoning. 

We initially applied our method in a small $L=62.5h^{-1}$ Mpc volume to explore how the bulk properties
of groups and poor clusters are affected by different feedback components, concentrating
on the \lxtx\ relation. The noteworthy results of this preliminary investigation are:

\begin{enumerate}
\item The star formation and supernova feedback scheme employed in the Munich L-Galaxies SAM
  has a negligible effect on the entropy of the ICM, leading to an \lxtx\ relation that
  resembles the relation from a simulation with gravitational heating only. By contrast,
  the mean relation obtained from hydrodynamical simulations with radiative cooling and stellar
  feedback typically lies closer to the observed one, particularly on cluster
  scales. However, these simulations tend to over-produce stars, whereas the fraction of baryons
  locked-up in stars within the virial radius of our clusters is, on average, approximately $9\%$, in excellent agreement wth observational estimates.

\item Incorporating a simple model for quasar mode AGN feedback in a simulation with
  L-Galaxies stellar feedback leads to X-ray luminosities considerably in excess of
  observed values, if the thermal coupling efficiency is $\epsilon_\mathrm{f}=0.05$. The
  choice $\epsilon_\mathrm{f}=0.05$ ensures the local black hole-bulge mass relation is
  recovered in hydrodynamical simulations including models for black hole growth and
  associated quasar activity \citep{DSH05}. However, our results suggest that
  simulations which employ the quasar mode feedback scheme of \citet{DCS08} would be
  unable to explain the observed scaling of X-ray luminosity with temperature. 

We find that reproducing the desired steepening of the \lxtx\ relation on group scales actually requires
  a much larger coupling efficiency: $\epsilon_\mathrm{f}=0.4$. In this case the hot gas
  fractions of our simulated clusters also broadly agree with observations. This indicates that the balance between quasar and radio mode feedback needs to be adjusted in L-Galaxies if the model is to simultaneously account for the observed properties of galaxies and the ICM.

\item We have implemented the recent \citet{BMB08} AGN feedback model in a hybrid
  simulation with stellar feedback from L-Galaxies. In this model, radio mode feedback can
  eject X-ray emitting gas from central regions of a halo, reducing the gas density and
  thus X-ray luminosity. This feedback mechanism is more efficient in lower-mass systems,
  causing the \lxtx\ relation to become steeper on group scales as desired. Indeed, we
  find that the relation obtained from our simulation agrees well with observational data
  on the mass scales probed. This is also true of the hot gas fractions of our simulated
  objects, demonstrating that significant AGN heating is a key ingredient in shaping groups and clusters.

\end{enumerate}

Once we had established a model capable of reproducing the observed \lxtx\ relation and
gas fractions of groups and poor clusters, we investigated whether this best-fit model
could also explain the properties of richer clusters by performing a hybrid simulation in
a larger $L=125h^{-1}$ Mpc volume. Our results can be summarised as follows:

\begin{enumerate}
\item The observed \lxtx\ relation is successfully recovered on all mass scales, apart
  from the occasional object that has an anomalously high temperature due to a recent
  injection of energy from the central AGN. The variation in scatter along the relation
  also compares favourably with observations, although there appears to be a lack of objects at the upper-luminosity edge of the observed \lxtx\ relation since we do not form CC systems in our non-radiative simulations.

\item The gas fractions of our groups and clusters are in broad agreement with
  observational data, displaying a similar degree of scatter. We find that AGN feedback
  significantly lowers the hot gas fraction in groups and poor clusters. This is because
  low-mass systems have a shallow potential well and AGN heating can efficiently drive
  X-ray emitting gas from their central regions to their outskirts. By contrast, massive
  clusters retain a greater fraction of their gas (up to $70\%$ of the cosmic baryon
  fraction) since they have a much larger binding energy and AGN feedback cannot
  effectively expel gas from the halo.

\item The radial entropy profiles of our simulated clusters begin to flatten off at $r\approx r_{500}$, departing from the scaling $S\propto r^{1.1}$ observed in cluster outskirts. By contrast, observed entropy profiles typically flatten off at smaller radii $r\approx 0.2r_{200}$ (e.g \citealt{PSF03,PAP06}). Consequently, we tend to over-estimate the core entropy. The core temperature is also higher than expected from observations of similar mass objects. However, profiles can only be reliably measured for X-ray bright objects, which are probably a biased sample of the population of groups and poor clusters. Our model may provide a reasonable description of X-ray under-luminous systems for which is it not possible to extract profiles. Nevertheless, if we are to reproduce the observed profiles of the majority of X-ray bright groups and poor clusters, the core entropy and temperature must be reduced, potentially by including radiative cooling in our simulations.
\end{enumerate}

The work described in this paper represents merely the first stage in the development of our method. Even so, we have obtained several encouraging results with this simple initial model. Following this success, we are currently resimulating a sample of rich clusters ($M_{500}\approx 10^{15}h^{-1}M_{\sun}$) extracted from the Millennium volume with our best-fit feedback model. In such massive objects the gas cooling time is longer, so the lack of radiative cooling in our simulations is less of an issue. Indeed, preliminary results suggest that the density, temperature and entropy profiles we obtain are in better agreement with observed profiles. These results will be presented elsewhere. 

Our primary goal in future work is to self-consistently incorporate radiative cooling into
our hybrid approach, rather than relying on the simple cooling recipes employed
in SAMs. These recipes usually assume that haloes have a
spherically-symmetric isothermal gas distribution but, in general, neither of
these assumptions will hold in hydrodynamical simulations. To circumvent this
problem, we intend to fully couple SAMs to radiative simulations, so that the gas distribution in the
simulation governs star formation, black hole growth and associated feedback in the
SAM. This is a non-trivial task, requiring the simulation and SAM to be run
simultaneously. Including gas cooling in our simulations should enable us
to produce more realistic ICM profiles in cluster cores, where the cooling time is short, since cooling
acts to lower the gas entropy in such regions.

Once our model is fully developed and tested, we shall conduct simulations in the full
Millennium volume. The idea is to generate a large sample of rich clusters that are
consistent with the high-quality X-ray data available on these scales. An example of an
important application of such a sample would be modelling the selection functions
of X-ray surveys (e.g. \citealt{SVL08}). This is essential to exploit the full power of
clusters as cosmological probes of the expansion history of the universe.

\acknowledgements

We thank V.~Springel for supplying the merger tree software and G.~De Lucia for providing
the code for the L-Galaxies semi-analytic model. We are also grateful to A.~Sanderson and A.~Vikhlinin for making their observational data available to us. All simulations were performed using the
Virgo Consortium Cosmology Machine at the Institute for Computational Cosmology,
Durham. This work was supported by a Science and Technology Facilities Council rolling
grant.

\bibliographystyle{apj} 
\bibliography{apj-jour,bibliography}

\end{document}